\documentclass[letterpaper, 10 pt, conference]{ieeeconf}  % Comment this line out if you need a4paper

\IEEEoverridecommandlockouts                              % This command is only needed if 
\usepackage{cite}
\usepackage{amsmath,amssymb,amsfonts}
\usepackage{algorithmic}
\usepackage{graphicx}
\usepackage{textcomp}
\usepackage{xcolor}
\usepackage{adjustbox}
\usepackage{booktabs}
\usepackage{hyperref}
\usepackage{epstopdf}
\usepackage{float}
\usepackage{hyperref}
\usepackage{tikz}

\title{\LARGE \bf
Context-Aware Vision Language Foundation Models for Ocular Disease Screening in Retinal Images
}

\author{Lucie Berger,$^{1}$ Mathieu Lamard,$^{2}$ Philippe Zhang,$^{3}$ Laurent Borderie,$^{3}$ \\
        Alexandre Le Guilcher,$^{3}$ Pascale Massin,$^{4}$ Béatrice Cochener,$^{5}$ Gwenolé Quellec,$^{6}$ and Sarah Matta$^{2}$% 
\thanks{$^{1}$ L. Berger is with IMT Atlantique, Brest, F-29200, France.}%
\thanks{$^{2}$ M. Lamard and  S. Matta are with Université de Bretagne Occidentale, Brest, F-29200 France and Inserm, UMR 1101, Brest, F-29200 France. {\tt\small sarah.matta@univ-brest.fr}}%
\thanks{$^{3}$ L. Borderie and A. Le Guilcher are with Evolucare Technologies, Villers-Bretonneux, F-80800 France.}%
\thanks{$^{4}$ P. Massin is with Service d’Ophtalmologie, Hôpital Lariboisière, APHP, Paris, F-75475 France.}%
\thanks{$^{5}$ B. Cochener is with Inserm, UMR 1101, Brest, F-29200 France, Université de Bretagne Occidentale, Brest, F-29200 France and Service d'Ophtalmologie, CHRU Brest, Brest, F-29200 France.}%
\thanks{$^{6}$ G. Quellec is with Inserm, UMR 1101, Brest, F-29200 France}%
\thanks{This work was supported by the French National Research Agency under the LabCom program (ANR-19-LCV2-0005 - ADMIRE project).}
}
\newcommand\submittedtext{%
  \footnotesize This work has been submitted to the IEEE for possible publication. Copyright may be transferred without notice, after which this version may no longer be accessible.}

\newcommand\submittednotice{%
\begin{tikzpicture}[remember picture,overlay]
\node[anchor=south,yshift=10pt] at (current page.south) {\parbox{\dimexpr0.65\textwidth-\fboxsep-\fboxrule\relax}{\submittedtext}};
\end{tikzpicture}%
}

\begin{document}

\maketitle
\thispagestyle{empty}
\pagestyle{empty}
%%%%%%%%%%%%%%%%%%%%%%%%%%%%%%%%%%%%%%%%%%%%%%%%%%%%%%%%%%%%%%%%%%%%
\begin{abstract}
Foundation models are large-scale versatile systems trained on vast quantities of diverse data to learn generalizable representations. Their adaptability with minimal fine-tuning makes them particularly promising for medical imaging, where data variability and domain shifts are major challenges. Currently, two types of foundation models dominate the literature: self-supervised models and more recent vision-language models. In this study, we advance the application of vision-language foundation (VLF) models for ocular disease screening using the OPHDIAT dataset, which includes nearly 700,000 fundus photographs from a French diabetic retinopathy (DR) screening network. This dataset provides extensive clinical data (patient-specific information such as diabetic health conditions, and treatments), labeled diagnostics, ophthalmologists text-based findings, and  multiple retinal images for each examination. Building on the FLAIR model—a VLF model for retinal pathology classification —we propose novel context-aware VLF models (e.g jointly analyzing multiple images from the same visit or taking advantage of past diagnoses and contextual data) to fully leverage the richness of the OPHDIAT dataset and enhance robustness to domain shifts. Our approaches were evaluated on both in-domain (a testing subset of OPHDIAT) and out-of-domain data (public datasets) to assess their generalization performance. Our model demonstrated improved in-domain performance for DR grading, achieving an area under the curve (AUC) ranging from 0.851 to 0.9999, and generalized well to ocular disease detection on out-of-domain data (AUC: 0.631-0.913).

\end{abstract}

\submittednotice

%%%%%%%%%%%%%%%%%%%%%%%%%%%%%%%%%%%%%%%%%%%%%%%%%%%%%%%%%%%%%%%%%%%%%%%%%%%%%%%%
\section{INTRODUCTION}

Artificial intelligence has revolutionized medical diagnostics, particularly in ophthalmology, where automated systems have excelled in detecting conditions such as diabetic retinopathy (DR) and age-related macular degeneration (AMD) \cite{matta2022automatic}. However, these advances are often constrained by the limitations of task-specific models (TSM), which rely on  extensive labeled datasets, are computationally intensive to train, and often exhibit poor generalization when applied to new populations, pathologies, or imaging conditions \cite{matta2024systematic}. Domain shifts—variations in data acquisition protocols, device types, or patient demographics—pose a persistent challenge, leading to significant performance drops. In response to these challenges, foundation models have emerged as a versatile solution \cite{chia2024foundation}. Unlike TSM, foundation models are trained on large, diverse datasets to learn generalizable representations that can be fine-tuned for specific downstream tasks. This adaptability reduces reliance on large labeled datasets, making them particularly valuable in contexts such as rare diseases or underrepresented populations.

Early foundation models leveraged self-supervised learning techniques \cite{huang2023self}, including generative approaches like the RETFound model \cite{zhou2023foundation} and contrastive methods like SimCLR \cite{chen2020simple}, to create robust image representations without manual annotations, useful in a field where annotating large datasets is both labor-intensive and expensive. More recently, vision language foundation (VLF) models have emerged, combining both visual and textual inputs to create richer, cross-modal representations. These models are particularly effective in zero-shot classification, enabling predictions for tasks they were not trained for \cite{shu2022test}. One notable example of VLF models is the FLAIR model \cite{silva2025foundation}, which has been trained  and validated on an extensive collection of 38 datasets, with 288,307 images and 101 different target categories. Its pretraining objective was to align image-text representations using contrastive learning. FLAIR has been evaluated for zero-shot classification and adaptation to downstream tasks and domains. Despite its effectiveness, FLAIR has limitations: the textual data used for training is primarily derived from diagnostic labels, and lacks  comprehensive patient-specific details. This constraint limits the richness of its encoder's representations.

Building on FLAIR's architecture, this study aims to develop an enhanced model that integrates clinical context and patient history by leveraging the OPHDIAT dataset, a large dataset collected from a diabetic population screened for DR. Beyond providing patient-specific clinical information, the OPHDIAT dataset \cite{massin2008ophdiat} organizes images by examinations, with each eye linked to a diagnosis and each examination accompanied by an ophthalmologist’s overall conclusion. This structure offers labeled data and free-text descriptions for each image, as well as longitudinal data from previous examinations and valuable context for each patient. 

In this work, we developed several context-aware strategies including associating labels with individual images, associating labels with several images - either from one eye or from both eyes - collected during one examination. Finally, we investigated the impact of integrating clinical data and results from previous examinations to further improve the robustness and generalization of the model.

\section{Background - FLAIR}
FLAIR is a multimodal model designed to learn representations from image-text pairs in the domain of retinal imaging. Its architecture comprises a vision encoder  based on ResNet-50 and a language encoder utilizing BioClinicalBERT. The model is trained using contrastive objectives to align images and their corresponding textual descriptions.
During pretraining, FLAIR aims to learn feature representations that minimize the distance between matched image-text pairs while maximizing the distance between unmatched pairs. To achieve this, image-text pairs were constructed based on the categorical label information. For each pair, a scalar product is computed between their feature vectors and stored in a matrix, where each element (i,j) represents the score for the pair (text$_{i}$, image$_{j}$). 
This approach encourages samples belonging to the same category to have closely aligned feature representations \cite{silva2025foundation}.

Given the scarcity of publicly available datasets with accompanying textual information, FLAIR generates text descriptions using categorical labels in the template:  ``a fundus image of [label]''. These labels are further enriched (i.e., augmented) with domain-specific knowledge. For instance, the label “myopia” is described with terms like  “tilted disc, peripapillary
atrophy, macular atrophy”. 
The pretrained model was evaluated through zero-shot inference on unseen categories from three public datasets: MESSIDOR (France) \cite{decenciere2014feedback}, REFUGE (China) \cite{orlando2020refuge}, and FIVES (China) \cite{jin2022fives}.

\section{OPHDIAT dataset}
%\subsubsection{OPHDIAT dataset description}
The OPHDIAT dataset \cite{massin2008ophdiat} originates from the OPHDIAT multi-center screening network developed by the Public Assistance Hospitals of Paris for DR screening. It comprises 164,659 examinations grouping, 763,849 images from 101,383 patients collected between 2004 and 2017. 
OPHDIAT’s components can be
presented as follows:
\begin{itemize}
    \item Images: Each screening examination comprised multiple retinal images, including optic nerve head-centered and macula-centered images of both eyes.
    \item Diagnosis: each eye was graded for DR by certified ophthalmologists according to the International Clinical Diabetic Retinopathy Scale \cite{wilkinson2003proposed}. The severity levels included: no diabetic retinopathy (noDR), mild diabetic retinopathy (mildDR), moderate diabetic retinopathy (modDR), severe diabetic retinopathy (sevDR), proliferative diabetic retinopathy (prolDR) and high-risk proliferative diabetic retinopathy (HR-prolDR). The presence of additional pathologies, such as diabetic macular edema (DME), glaucoma (G), cataract, hypertensive retinopathy, and macular dystrophy, was assessed on a binary scale. The presence of other pathologies was indicated in a text.
    \item Ophthalmologist's Conclusion: findings for the examination were documented by ophthalmologists in free-form text.
    \item Image Quality: Evaluation of the quality of the retinal images for each eye examination.
    \item Clinical Data: Patient-specific information, including physical attributes (e.g., height, weight, gender) and health conditions (e.g., diabetes type, year of diabetes diagnosis,  hypertension, treatment, cholesterol, creatinine).
    \item Metadata: General information such as the center, device, technician, ophthalmologist, and the date of the examination (providing temporal context for longitudinal studies).
\end{itemize}
Inspired by FLAIR, the OPHDIAT dataset will serve us for developing context-aware VLF models.

\section{Methodology}
In this section, first, we will present our partitioning and preprocessing of the  OPHDIAT dataset. Then, we will present our context-aware VLF methods, and our evaluation methodology.

\subsection{OPHDIAT partitioning and preprocessing }
For the purpose of this study, eye laterality was determined for each image using our laterality classification algorithm \cite{matta2023towards}.
For quality assurance, a subset comprising 21,054 images from 4,850 patients was annotated by two human readers. We utilized this subset of images with higher-quality annotations as a testing set. Then, we ensured that these patients are not present in the development of the algorithms. The remaining data (i.e., the development set) was divided into 80$\%$ training and 20$\%$ validation, ensuring no overlap of data from the same patient across splits. Only gradable images were considered for the validation subset. A total of 564,186 images were used for training and 127,336 images for validating the models.  

The OPHIDAT dataset contains French tabular and textual data. For the purpose of preprocessing the dataset, the   Gemma 2 instruct (gemma-2-27b-it-GGUF) model \cite{team2024gemma} was employed. It was prompted to generate 3  summaries while translating from French to English: 1) \textit{clinical context examination summary}: a summary of all tabular data present in the clinical data and the metadata components; 2) \textit{left eye diagnosis summmary:} a summary of all tabular data present in the diagnosis component related to the  examination of the left eye and of the ophthalmologist's conclusion; 3) \textit{right eye diagnosis summmary:} a summary of all tabular data present in the diagnosis component related to the examination the right eye  and of the ophthalmologist's conclusion.  In addition, it was prompted to translate the ophthalmologist's conclusion (denoted as Concl) from French to English.

\subsection{ Context-Aware VLF Models Development}
Our approach involves developing context-aware models to process various inputs, including images (number of images per examination) and text (generated from diagnostic labels or ophthalmologist's conclusion, etc.). First,  we present Base VLF models tailored to different input types. To enhance performance, we also develop  combined VLF models that integrate multiple Base models.  Furthermore, we propose a Clinical-Temporal VLF model, which incorporates clinical data and patient's historical examination results.

\subsubsection{Base VLF Models}

Four context-aware base VLF models were developed, each designed to handle different input image configurations and associated diagnostic information. \begin{itemize}
    \item Unilateral-S: Processes a single image per examination, associating it with a labeled diagnostic.
    \item Unilateral-D: Handles two retinal images (one retina-centered and one macula-centered eye)  from a single eye examination. Two models were developed: one  that associate an eye examination with pathology labels, and another that employs Gemma eye-diagnosis generated summaries, named Unilateral-D (eye diagnosis summary).
    \item Bilateral: Aggregates four images (one retina-centered and one macula-centered image from both eyes) per examination and aligns them with the ophthalmologist’s overall conclusion (Concl).
\end{itemize}

The FLAIR architecture was modified to accommodate different inputs. Regarding image modality, the Root Mean Square (RMS) was applied after the projection layer to combine features of multi-image input into a single vector.
Regarding text modality, different strategies were explored. For integrating labeled diagnostics, text was generated following a template that incorporated image quality, such as  “a [quality] fundus image of [label]”. Labels were augmentated as in FLAIR. Another strategy consisted of training the model on the ophthalmologist's conclusions (Concl). 

\subsubsection{Combined VLF Models}
%Training strategies mainly focused on two types of textual inputs: structured sentences derived from labeled diagnostics and free-text conclusions translated from ophthalmologist notes. 
To enhance model performance and enrich input representation, we explored the combination of different VLF Base models. This was achieved by calculating individual losses for each model and subsequently summing them prior to the backpropagation process.

\subsubsection{Clinical-Temporal VLF Model} Integrates clinical and temporal (past examination) data  by combining clinical information with the diagnosis from the last exam using the template  \textit{``[clinical data]. The previous exam showed [result of the last exam]"} for patients with prior exams, and  \textit{``[clinical data]. It is the first exam."} for first-time examinations. Here, ``clinical data" refers to clinical context examination summary generated by Gemma and the ``result of the last exam" denotes the diagnostic outcome from the previous examination. 

\begin{figure}[ht]
\centering
    \includegraphics[width=\linewidth, height=5cm]{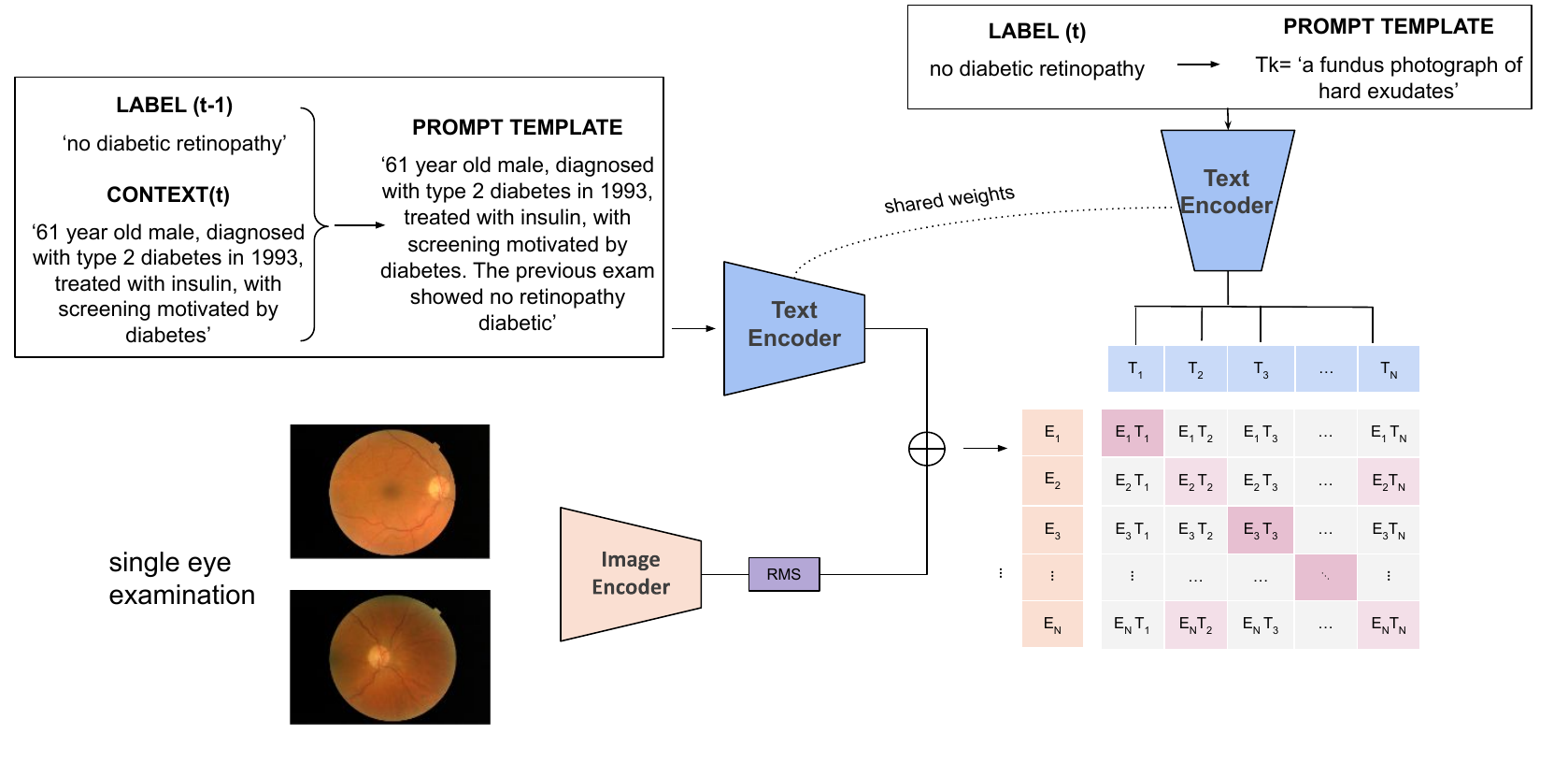}
    \caption{Framework overview for training with clinical data and prior examination result included}
    \label{fig:Fig1}
\end{figure}

This textual information is then encoded  through a text encoder to create an embedding that captures both the clinical and temporal data (see Fig.\ref{fig:Fig1}).
To form a multi-modal representation, this generated embedding is combined with the image feature vector from the current examination by summing them. The same textual encoder is used to compute feature vector for the labels in the current examination, and these encoded features are aligned with the fused image-text feature vector using a contrastive approach. To enhance adaptability to standard datasets lacking clinical data or longitudinal examinations, these data were incorporated into only half of the images during training.

\subsection{Implementation Details} 

Each model was trained and validated using the OPHDIAT development set. The final model weights were selected based on performance on the OPHDIAT validation subset, optimizing for area under the receiver characteristic curve (AUC) metric for the DR zero-shot grading task. 
The models were trained for 15 epochs using the AdamW optimizer with a base learning rate of $10^{-4}$. A warm-up cosine scheduler was used during the first dataset pass. 

\subsection{Models Evaluation}
Models were evaluated on the basis of their zero-shot classification performance using AUC metric on the OPHDIAT testing subset for DR grading. The performances of these models were compared with the original FLAIR model. In addition, we assessed the zero-shot classification generalization performance for detecting  diseases under domain shifts using the public datasets MESSIDOR (DR, DME), REFUGE (glaucoma), and FIVES (Normal, DR, glaucoma, AMD). 

\section{Results and Discussions}

\begin{table*}[!t]
\caption{Performance of context-aware VLF models and baseline method on in-domain and out-of-domain data}
 \label{tab:tab_1}
 \begin{tabular}{p{4.4cm}|c c c c c c||c c c}
        \toprule
  \textbf{Model} & \multicolumn{6}{c||}{\textbf{OPHDIAT }} & \textbf{MESSIDOR} & \textbf{FIVES} & \textbf{REFUGE} \\ 
 \multicolumn{1}{c|}{\textbf{}} & \textbf{noDR} & \textbf{mildDR} & \textbf{modDR} & \textbf{sevDR} & \textbf{prolDR} & \textbf{HR-prolDR} &  &   &  \\ 
        \midrule
        FLAIR & 0.73 & 0.58 & 0.814 & 0.877 & 0.705 & 0.93 &0.888 &0.903 & \textbf{0.918} \\
        \midrule
        \multicolumn{10}{c}{\textbf{Base VLF Models}} \\ 
        \midrule
        Unilateral-S  & 0.852 & 0.676 & 0.89 & 0.985 & 0.977 &0.995  & 0.865 & 0.957& 0.838\\
        Unilateral-D  & \textbf{0.915} & 0.828 & \textbf{0.97} & 0.993 & \textbf{0.992} & 0.99 & 0.9059& 0.955 & 0.810\\
        Unilateral-D (Eye diagnosis summary) & 0.898 & 0.82 & 0.969 & \textbf{0.996} & 0.98 & 0.998& 0.895& 0.364 &0.199\\
        Bilateral (Concl) & 0.893 & 0.743 & 0.951 & 0.992 & 0.983 & 0.987 &0.882 &0.964 &0.768 \\
        \midrule
        \multicolumn{10}{c}{\textbf{Combined VLF Models}} \\ 
        \midrule
        Bilateral (Concl) + Unilateral-S  &0.91 & 0.819 & 0.93 & 0.989 & 0.978 & \textbf{0.999} &0.9079 &0.961 &0.894 \\
        Bilateral (Concl) + Unilateral-D  & 0.896 & 0.809 & 0.965 & 0.993 & 0.988 &\textbf{0.999}& 0.8999 &0.939 & 0.863\\
        Bilateral (Concl) + Unilateral-D + Unilateral-S & 0.893 & 0.743 & 0.951 & 0.992 & 0.983 & 0.987 & 0.9035& \textbf{0.965} &0.810 \\
        \midrule
        \multicolumn{10}{c}{\textbf{Clinical-Temporal VLF Model}} \\ 
        \midrule
        Unilateral-D Clinical-Temporal model & \textbf{0.915} & \textbf{0.851} & \textbf{0.97} & 0.994 & 0.983 & \textbf{0.999} & \textbf{0.913}& 0.792&0.631\\
        \bottomrule
    \end{tabular}
\end{table*}

In this work, we developed context-aware VLF models using the OPHDIAT development set. Table \ref{tab:tab_1} reports zero-shot classification AUC results obtained on the OPHDIAT test subset and on the out-of-domain data.

For DR grading on the OPHDIAT dataset, the context-aware VLF models achieved good performances with AUC values ranging from 0.676 to 0.999. The best-performing model incorporated clinical data and prior examinations, highlighting that integrating comprehensive clinical information and longitudinal history can significantly enhance clinical performance. The results also demonstrate that combining multiple models improves overall performance on in-domain data. Furthermore, the context-aware VLF  models outperformed the FLAIR model on the OPHDIAT testing subset.

On public datasets, AUC generalization across context-aware VLF models (excluding Unilateral-D (eye diagnosis summary)) ranged from 0.631 and 0.95. The Unilateral-D (eye diagnosis summary) exhibited poorer generalization, likely due to the variability in text generated by Gemma. The best model for zero-shot testing under domain shift conditions employed a hybrid approach that combined a model trained on full examinations with ophthalmologist conclusions and a model trained on single-image labels. However, the model trained using contextual data produced more variable results when tested on public datasets, likely because these datasets lacked the detailed information available in OPHDIAT.

The FLAIR model generally performed better on REFUGE. This discrepancy can be due to our selection method for the optimal model, which was based on its DR performance on the validation set. 

\section{Conclusion}
In conclusion, our proposed context-aware VLF models have shown promising potential for screening ocular pathologies in fundus photographs.  To further enhance the performance of the model, several avenues can be explored. These include experimenting with alternative fusion strategies or fine-tuning the trained models for downstream tasks using few-shot classification approaches.

\bibliographystyle{IEEEtran}
\bibliography{biblio.bib}

\end{document}